\documentclass[%
 reprint,
 amsmath,amssymb,
 aps,
]{revtex4-2}

\usepackage{graphicx}
\usepackage{dcolumn}
\usepackage{bm}

\begin{document}

\preprint{APS/123-QED}

\title{Ultra-bright Quantum Photon Sources on Chip}


\author{Zhaohui Ma}
\altaffiliation{These authors contributed equally}
\author{Jia-Yang Chen}
\altaffiliation{These authors contributed equally}
\author{Zhan Li, Chao Tang, Yong Meng Sua, Heng Fan}
\author{Yu-Ping Huang}%
 \email{yhuang5@stevens.edu}
\affiliation{%
 Department of Physics, Stevens Institute of Technology, 1 Castle Point Terrace, Hoboken, NJ 07030, USA
}%
\affiliation{
 Center for Quantum Science and Engineering, Stevens Institute of Technology, 1 Castle Point Terrace, Hoboken, NJ 07030, USA
}%

\date{\today}

\begin{abstract}
Quantum photon sources of high rate, brightness, and purity are increasingly desirable as quantum information systems are quickly scaled up and applied to many fields. Using a periodically poled lithium niobate microresonator on chip, we demonstrate photon-pair generation at high rates of 8.5 MHz and 36.3 MHz using only 3.4-$\mu$W and 13.4-$\mu$W pump power, respectively, marking orders of magnitude improvement over the state-of-the-art. The measured coincidence to accidental ratio is well above 100 at those high rates and reaches $14,682\pm 4427$ at a lower pump power. The same chip enables heralded single-photon generation at tens of megahertz rates, each with low auto-correlation $g^{(2)}_{H}(0)=0.008$ and $0.097$ for the microwatt pumps. Such distinct performance, facilitated by the chip device's noiseless and giant optical nonlinearity, will contribute to the forthcoming pervasive adoption of quantum optical information technologies.    

\end{abstract}

\maketitle

Efficient and noiseless parametric processes in photonic integrated circuits are enabling resources for a breadth of value-creation classical and quantum optical applications. With high quality factor, small mode volume, and agile dispersion engineering, various integrated photonic resonators, such as those of silicon (Si) \cite{jiang2015silicon, lu2016heralding, Ma:17}, gallium phosphide (GaP) \cite{logan2018400}, gallium arsenide (GaAs) \cite{chang2019strong}, silicon nitride (Si$_3$N$_4$) \cite{lu2019chip}, aluminum nitride (AlN) \cite{bruch201817}, and lithium niobate (LN) \cite{chen2019ultra,lu2019periodically}, have demonstrated nonlinear frequency conversion with higher performance than their waveguide-based counterparts. Among them, thin-film lithium niobate on insulator (LNOI) is being intensively investigated due to its exceptional linear optical properties, outstanding electro-optical responses, and strong second-order ($\chi^{(2)}$) nonlinearities. Thus far, a variety of LNOI microcavities have been demonstrated with impressive optical nonlinearities \cite{chen2019ultra,lu2019periodically, hao2020second, lu2020towards}.

Compared with those utilizing third-order ($\chi^{(3)}$) nonlinearities, $\chi^{(2)}$-based devices enjoy orders of magnitude stronger nonlinear optical effects while not subjet to undesirable side $\chi^{(3)}$  processes. They are thus advantageous for quantum photonics applications like photon-pair generation \cite{kwiat1995new}, heralded single-photon creation \cite{fasel2004high, kaneda2019high}, and quantum frequency conversion \cite{zaske2012visible, Huang:13}. Nevertheless, existing $\chi^{(2)}$ integrated circuits for quantum applications \cite{moore2016efficient, luo2017chip, guo2017parametric} have yet to reach their full potential, hindered by suboptimal nonlinear susceptibilities (e.g., $d_{31}\sim$4.7 pm/V in LN \cite{moore2016efficient} and 4--6 pm/V in AIN \cite{guo2017parametric}), poor mode overlap \cite{luo2017chip}, phase mismatching, etc.

Here, we present the first quantum circuit in LNOI that overcomes all shortcomings and approaches the ideal performance promised by the material. It is a triply resonant, Z-cut periodically-poled lithium niobate (PPLN) microring resonator, where the light waves are all in the low-loss fundamental modes, with nearly perfect overlap, and interact through LN's largest nonlinear susceptibility tensor element ($d_{33}\sim$27 pm/V). Such an all-optimized device leads to spontaneous parametric donwconversion (SPDC) at an unprecedented efficiency, by which we demonstrate ultra-bright photon pair generation at a 2.7 MHz rate per microwatt of pump power. This ultrahigh efficiency corresponds to 100 times or more improvement over the state of the art across all existing material platforms (see Table \ref{table1}). Meanwhile, the measured coincidence to accidental ratio of the photon pairs reaches $14,682\pm 4427$, which notes the ultrahigh purity of the photon-pair states with perfect correlation, and with nearly zero noise. In addition, with the same device we demonstrate heralded generation of single photons, observing a 8.5 MHz rate and 0.008 auto-correlation. Our work presents a nearly ideal LNOI circuit to deliver its promises in quantum nonlinear optics, and marks a new performance milestone in quantum light sources for various applications. As many other circuits can be incorporated on LNOI, a series of functional quantum chip developments are expected to follow in the near future.   

\textbf{Device Design and Fabrication.} In a single-mode cavity, the effective Hamiltonian describing  non-degenerate spontaneous parametric downconversion is:  
\begin{equation}
\hat{H}_\textbf{eff}  = \hbar\mathnormal{g}(\hat{a}_s\hat{a}_i\hat{b}_p^\dagger+\hat{a}_s^\dagger\hat{a}_i^\dagger\hat{b}_p), 
  \label{eq1}
\end{equation}
where $\{$$\hat{a}_s$, $\hat{a}_i$, and $\hat{a}_p$$\}$ are the annihilation operators for the signal, idler and pump photons, respectively, $\mathnormal{g}$ is the nonlinear coupling coefficient between pump and photon pairs, which is a function of each's mode volume, their spatial overlap, nonlinear susceptibility, and phase matching \cite{luo2019optical, Chen:19, guo2017parametric}. Assuming phase matched case and no cavity detuning, the total photon pair generation rate (PGR) is given by \cite{guo2017parametric}:
\begin{equation}
  \text{PGR} = \frac{64g^2}{\kappa_{s,t}+\kappa_{i,t}}\frac{\kappa_{p,e}}{\kappa^2_{p,t}} \frac{P_p}{\hbar\omega_p},
  \label{eq2}
\end{equation}
where $\kappa_{j,o} = \omega/Q_{j,o}$ are the cavity dissipation rates with $j=s,i,p$ for the signal, idler, and pump modes, respectively, and $o=e,t$ denoting external and total cavity dissipation rates. $Q_{j,0}$ is the quality factor of cavity mode. $\omega$ is the angular frequency and $P_p$ is the applied pump power. 

\begin{table*}
\caption{The state-of-the-art of photon-pair and heralded single-photon generation in $\chi^{(2)}$ and $\chi^{(3)}$ microcavities.}
\begin{ruledtabular}
\begin{tabular}{ccccccc}
Reference& Material Structure&On-chip pump power\footnote{$^{\dagger}$ is spontaneous four-wave mixing (SFWM) source, $^{\dagger\dagger}$ is Spontaneous parametric down conversion  (SPDC) source.}&PGR& PGR $@$ 1$\mu$W &CAR&$g_H^{(2)}$\\ \hline

Jiang \cite{jiang2015silicon}&Si $\mu$-disk&79 $\mu$W$^{\dagger}$ &855 kHz&137 Hz &274&$\cdot\cdot\cdot$\\
Lu \cite{lu2016heralding}&Si $\mu$-disk &12 $\mu$W$^{\dagger}$ &1.2 kHz&8.3 Hz&2610&0.003\\
Ma \cite{Ma:17} & Si $\mu$-ring &59 $\mu$W$^{\dagger}$&1.1 MHz&316 Hz &532&0.098\\
Ma \cite{Ma:17} & Si $\mu$-ring &7.4 $\mu$W$^{\dagger}$&16 kHz&292 Hz &12105&0.005\\
Lu \cite{lu2019chip}& Si$_3$N$_4$ $\mu$-ring &500 $\mu$W$^{\dagger}$ &1 MHz&4 Hz& 50 &$\cdot\cdot\cdot$ \\
Lu \cite{lu2019chip}& Si$_3$N$_4$ $\mu$-ring &46 $\mu$W$^{\dagger}$ &4.8 kHz&2.27 Hz& 2280 &$\cdot\cdot\cdot$ \\
Frank \cite{moore2016efficient}& LN $\mu$-disk &16.7 $\mu$W$^{\dagger\dagger}$ &450 kHz&26.95 kHz&6 &$\cdot\cdot\cdot$\\
Luo \cite{luo2017chip}& LN  $\mu$-disk &115 $\mu$W$^{\dagger\dagger}$ &0.5 Hz&0.004 Hz& 43 &$\cdot\cdot\cdot$\\
Guo \cite{guo2017parametric}& AlN $\mu$-ring &1.9 mW$^{\dagger\dagger}$ &11 MHz&5.8 kHz&$\cdot\cdot\cdot$ &0.088\\
This work& PPLN $\mu$-ring &3.4 $\mu$W$^{\dagger\dagger}$ &8.5 MHz&2.53 MHz &451&0.008 \\
This work & PPLN $\mu$-ring & 13.4 $\mu$W$^{\dagger\dagger}$ &36.3 MHz & 2.7 MHz & $>$100& 0.097 \\
\end{tabular}
\end{ruledtabular}
\label{table1}
\end{table*}

\begin{figure}[ht]
  \centering
\includegraphics[width=3.2in]{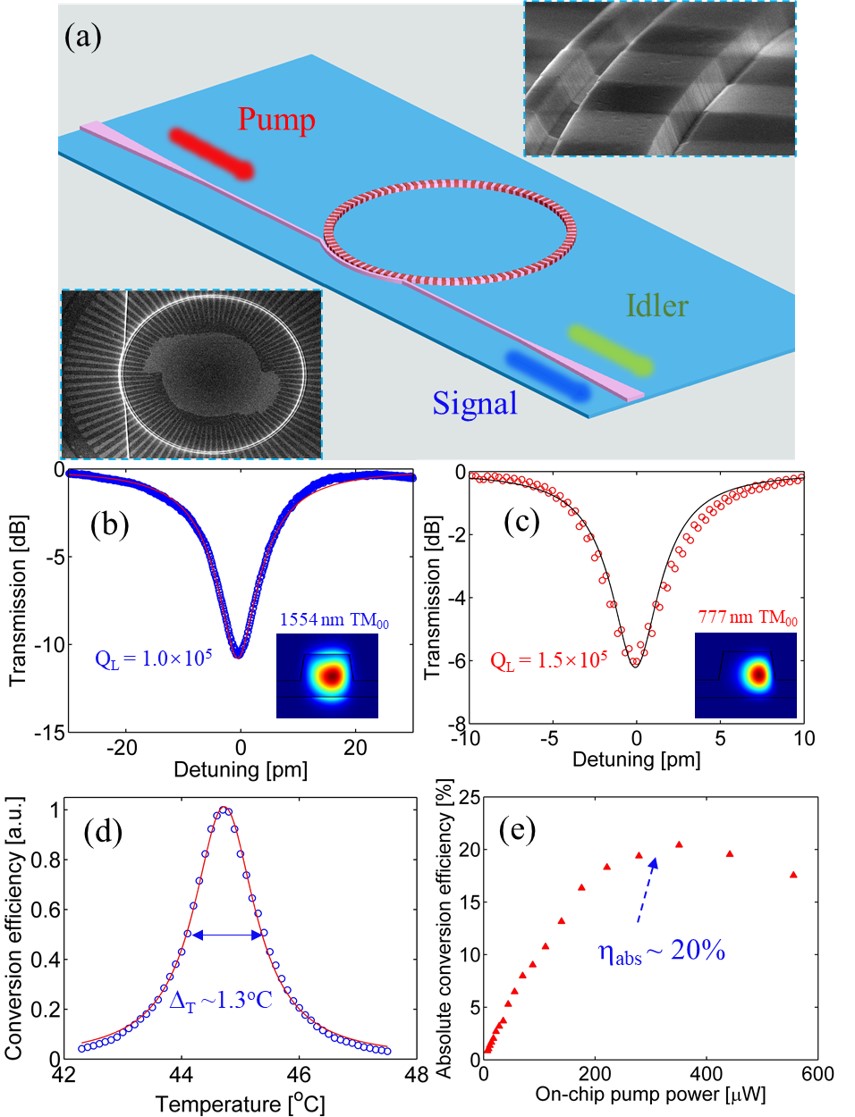}
\caption{Device design and characterization. (a) Schematic of the PPLN microring. Insets are the scanning electron microscopy images of the fabricated device and a zoom-in of the coupling region. (b) and (c) are the spectra of IR and visible quasi-TM$_{00}$ modes around 1554.6 nm and 777.3 nm, respectively. Insets are the simulated mode profiles. (d) Temperature dependency of SHG efficiency, with Lorentzian-fitted temperature bandwidth at 1.3 $^\circ$C. (e)  Absolute SHG efficiency ($P_{sh}/P_p$) versus the on-chip pump power.}
\label{fig1}
\end{figure}

The key to maximize the nonlinear coupling coefficient $g$ is quasi-phase matching. Here,  we apply concentric periodic-poling on a microring resonator to realize the strongest interaction between the fundamental quasi-transverse-magnetic (quasi-TM) cavity modes in infrared (IR) (signal and idler) and visible (pump) bands, while achieving ideal mode overlapping and utilizing the largest nonlinear tensor $d_{33}$ in LN. The schematic of our device is shown in Fig.~\ref{fig1} (a). The entire device is fabricated on a Z-cut LNOI wafer (NANOLN Inc.), with a 700-nm thick LN thin film bonded on 2-$\mu$m silicon dioxide layer above a silicon substrate. First, a concentric periodically poled region is created with a period of $3.85~\mu $m using a similar process described in \cite{chen2020efficient}. Then, standard electron beam lithography is used to define the microring structure with a top width of 1.6 $\mu$m and a radius of 55 $\mu$m. Finally, ion-milling is applied to shallowly etch the microring (430 nm ethed, 270 nm remained) with 70$^\circ$ sidewall angle. We obtain $Q_L (Q_0) \approx 1.0 (2.6)\times 10^5$ (IR mode) and $Q_L (Q_0) \approx 1.5 (2.3)\times 10^5$ (visible mode) as shown in Fig.~\ref{fig1} (b) and (c). A pulley waveguide (top-width 300 nm, gap 500 nm, see Fig.~\ref{fig1}(a)) is designed as the ring-bus waveguide coupler to attain simultaneously good coupling for both IR ($90\%$) and visible light ($75\%$). 

\textbf{Second-Harmonic Generation}. We first characterize the device nonlinearity using second-harmonic generation (SHG). Two polarized (quasi-TM) tunable lasers (Santec 550, 1500-1630 nm and Newport TLB-6712, 765-781 nm) and tapered fibers (OZ OPTICS) are used to independently characterize fiber-chip-fiber coupling, whose losses are measured to be 11.5 dB around 1554 nm and 17.5 dB around 777 nm. By sweeping the infrared pump laser and fine tuning the device's temperature, a quasi-phase matched SHG is achieved for optimal resonance modes at 1554.6 nm and its second-harmonic (SH) at 777.3 nm at 44.6 $^\circ C$. With 83 $\mu$W pump power in the input fiber, 80 nW SH power is collected in the output fiber. Accounting for the coupling loss, the normalized SHG efficiency $\eta_\mathbf{SHG} = P_{sh}/P^2_p$ is estimated to be $122,000 \%/W$, corresponding to single-photon nonlinearity $\eta_\mathbf{photon} = \eta_\mathbf{SHG} \hbar\omega^2_{p}/Q_{p,t} \approx 1.9\times10^{-6}$. Figure~\ref{fig1}(d) plots the temperature dependency of the SHG efficiency for the optimized cavity mode (around 1554.6 nm), showing a tolerance of 1.3$^\circ C$. The absolute SHG efficiency as a function of the pump power is shown in Fig.~\ref{fig1}(e), saturating at 20$\%$ with 350 $\mu$W on-chip pump power. This exceptional efficiency indicate good thermal stability and slightly over-coupled condition for the pump mode.

\begin{figure}[ht]
    \centering
    \includegraphics[width=0.48\textwidth]{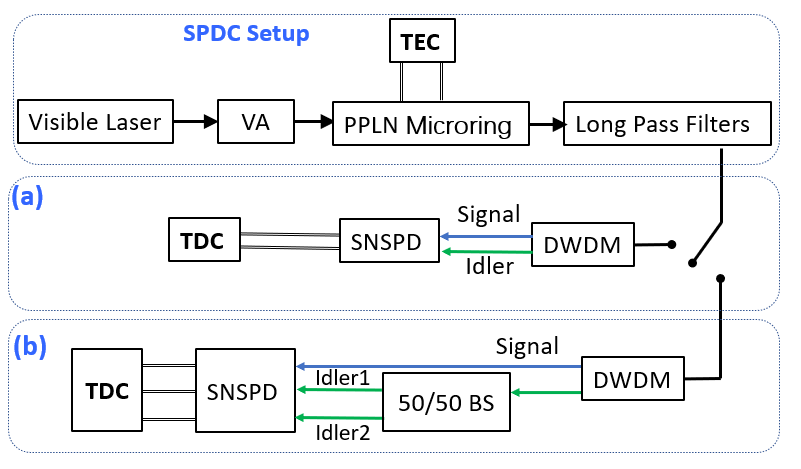}
    \caption{Experimental procedures for (a) photon-pair and (b) heralded single-photon generation. VA: variable attenuator; TEC: temperature electronic controller; DWDM: dense wavelength-division multiplexing; SNSPD: superconducting nanowire single-photon detectors; BS: beam splitter; TDC: time to digital converter.}
    \label{fig2}
\end{figure}

\textbf{Photon-Pair Generation.} As shown in Fig.\ref{fig2} (a), a continuous wave tunable pump laser (Newport TLB-6712,  765-781 nm) is connected to a visible fiber attenuator (OZ OPTICS) and a fiber polarization controller (FPC) for accurate control of the pump power and polarization state, respectively. The quasi-TM fundamental visible cavity mode at phase matched SHG wavelength (777.3 nm) is excited for optimum photon-pair generation. The generated photon pairs are coupled out via tapered fibers and passed through free-space long-pass filters with over 80 dB extinction ratio and 1.5 dB insertion loss (IL) for rejecting any residual visible pump. Cascaded Dense Wavelength Division Multiplexing (DWDM) filters with narrow full width at half maximum (FWHM) transmission bandwidth of 0.8 nm, are employed to demultiplex signal-idler photon pairs ($\sim$ 1.5 dB IL). To ensure the maximum detection efficiency, two FPCs ($\sim$ 1 dB IL) are used to independently optimize signal and idler photons before they enter superconducting single-photon detectors (4 channels SNSPDs, ID281, ID Quantique). The SNSPDs have dark-count rate of 50-100 Hz and detection efficiency of 85$\%$. They are connected to a time-tagging unit (ID 900) for photon counting and coincidence detection. Considering the timing jitter of the SNSPDs and timing resolution of the time tagger, the temporal resolution of the entire system in determining the arrival time of photon is about 46 ps (see Supplementary Materials). After taking all insertion loss, chip-fiber coupling loss and detection efficiency into account, signal and idler channel each has about 10 dB total loss. 

\begin{figure}[ht]
    \centering
    \includegraphics[width=0.48\textwidth]{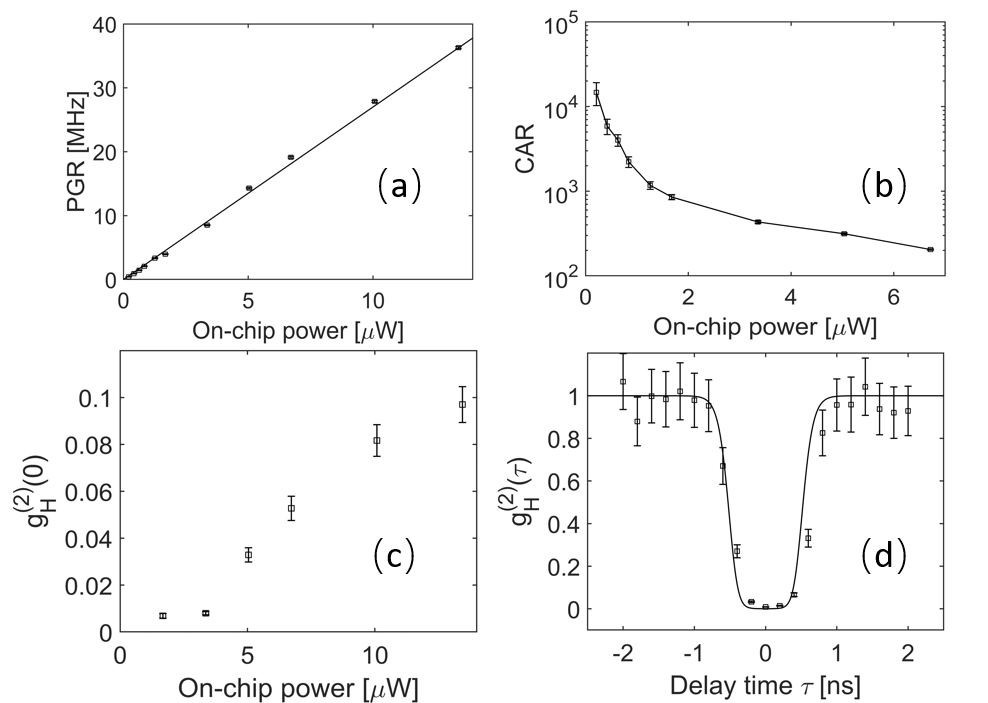}
    \caption{(a) Pair generation rate (PGR) and (b) coincidences-to-accidentals ratio (CAR)  and (c) auto-correlation function $ g_H^{(2)}(\tau)$  with zero delay time, as the function of on-chip pump power. (d) Heralded idler-idler correlation function $ g_H^{(2)}(\tau)$ with anti-bunching dip of $ g_H^{(2)}(0)=0.008\pm0.0007$ at 3.4 $\mu$W on-chip power. All error bars are estimated by Poissonian photon counting statistics.}
    \label{fig3}
\end{figure}

\textbf{Pair Generation Rate and Brightness} To extract the pair generation rate (PGR), we measure the two-photon coincidence ${N_{12}}$ and  single-photon detection rates of the signal (${N_1}$) and idler (${N_2}$), the latter at 1545.8 nm and 1563.5 nm, respectively. Then, $\text{PGR}$ is simply calculated as $N_1 N_2/N_{12}$. The result is presented in Fig.\ref{fig3} (a) for various pump power levels. The Brightness (B) of the photon-pair source, defined as PGR over on-chip pump power ($\text{B}=\text{PGR}/P_p$), is measured to be 2.7 MHz $\mu W^{-1}$, as inferred from the fitted slope of the data in Fig.\ref{fig3} (a). The observed brightness is about 2.6 times lower than the theoretical prediction of 7.0 MHz $\mu W^{-1}$, mainly due to its extreme sensitivity in the slight deviation from the ideal triply resonant condition. As shown in Table \ref{table1}, the brightness and PGR of our PPLN microring photon pair source are several orders magnitude higher compared to the previously demonstrated state-of-the-art microcavity based photon pair sources. The unmatched brightness and PGR observed in this work is facilitated by the giant single photon nonlinearity ($\sim$ $10^{-6}$), an excellent single-mode condition, and the pure SPDC process in this system. 

\textbf{Coincidence-to-Accidental Ratio (CAR)} We characterize the figure-of-merit of this PPLN microring photon pair source by measuring the CAR to quantify the multi-photon noise. Figure \ref{fig3} (b) shows the CAR measurements at various on-chip pump power levels. The CAR is calculated as $(N_{CO}/N_{AC})-1$, where $N_{CO}$ and $N_{AC}$ are defined as coincidence and accidental coincidence counts, respectively. Considering the cavity lifetime ($\tau_{c}\approx$ 80 ps), we choose the coincidence window to be 200 ps to cover the majority of the coincidence events.
The highest observed CAR is 14682$\pm$4427 with raw coincidence and accidental counts, i.e., without accidental subtraction, measured with 210 nW on-chip pump power. As the measured single photon detection rate is much higher than the dark-count rate of the detectors and excellent system timing resolution of about 46 ps, the CAR are primarily limited by multi-photon emission. Evidently, CAR decreasing exponentially with increasing pump power on chip, a well-known trade-off between a high mean photon probability and unwanted multi-photon noise of any probabilistic photon source. Nevertheless, a high CAR of $205\pm3$ can be achieved at detected photon flux of 0.9 MHz. The photon pair generation occurs simultaneously over multiple wavelength channels---as permitted by the cavity resonance---for the signal and idler pairs, with similarly high performance. To examine it, we have measured the CAR over five more channel pairs (see Supplementary Materials). At a on-chip pump power of 1.7 $\mu W$, the CAR values for those channels vary from 500 to 1000, indicating efficient generation of high quality photon pairs simultaneously over many wavelength channels. 

\textbf{Heralded Single-Photon Generation} Next we demonstrate heralded generation of single photons using the same chip. As shown in Fig.\ref{fig2} (b), We use the signal photons at 1545.8 nm for heralding single photons at the paired idler wavelength of 1563.5 nm. The idler photons are split by a 50/50 splitter ($\sim$ 1 dB IL) to Idler 1 and Idler 2 beams and their heralded auto-correlation $g_H^{(2)}(\tau)$ measurement are measured with $\tau$, the relative delay between the two beams. Then, the single photon counts, $N_{1,2,3}$ for signal, idler 1 and idler 2, respectively, double coincident counts $N_{12}$ and $N_{13}$ for signal \& Idler 1 and signal \& Idler 2, and triple coincident counts $N_{123}$ among all, are measured using the four-channel SNSPD. The timing window is 200 ps for all measurements, and each total counting period is 300 seconds. $g_H^{(2)}(\tau)$ is calculated as $N_{123} N_1/N_{12} N_{13}$ and then normalized. The results are shown in Fig.~\ref{fig3} (c), where the measured $g_H^{(2)}(\tau=0)$ is plotted as a function of the on-chip pump power. As shown, the best auto-correlation value at zero delay is measured to be $g_H^{(2)}(0) = 0.007\pm0.0016$, with the uncertainty estimated by the Poissonian photon counting statistics) in Fig.~\ref{fig3} (c) with detected count rate of $N_1$ = 250 kHz. Such a low auto-correlation points to the high fidelity with very low noise of the heralded single photon state. At a higher pump power, the auto-correlation remains at $ g_H^{(2)}=0.097\pm0.0076$ even when the detected photon rate of $N_1$ = 2 MHz. Figure \ref{fig3} (d) shows the idler-idler auto-correlation as a function of $\tau$, where a distinct anti-bunching dip is observed, where only one idler photon is detected upon one heralding event. Similarly, another non-degenerate pairs (signal-1551.7 nm and idler-1557.6 nm) (see Supplementary Materials) also recorded impressively low $g^2_H(0) = 0.005 \pm 0.0009$ with detected count rate of $N_1$ = 225 kHz, indicating the potential of the current PPLN microrings as a versatile single and paired photon sources for a variety of quantum optical information technologies.

In conclusion, we have demonstrated photon-pair and heralded single-photon generation of high rate and quality, in a Z-cut PPLN microring resonator. Enabled by giant single-photon nonlinearity, we have observed orders of magnitude improvement in efficiency over all existing integrated nonlinear realizations. With only microwatt pump power, the photon production reaches tens of megahertz. Thanks to the nearly perfect single mode condition and absence of background optical processes, those photons are in pure quantum states. Also, we show that such high-quality single and correlated photons are created over multiple wavelength channels simultaneously, lending our devices to quantum information applications such as high-dimensional entangled quantum states \cite{kues2017chip} and optical quantum logic \cite{imany2019high}. Finally, the demonstrated quantum circuits are ready to be integrated with other optical elements, such as electro-optical modulators and frequency converters, on the same LNOI chip, by which high-speed, reconfigurable, and multifaceted functional quantum devices can be realized \cite{jin2019high} . 

\begin{acknowledgments}
The research was supported in part by National Science Foundation (Award \#1641094 \& \#1842680) and National Aeronautics and Space Administration (Grant Number 80NSSC19K1618). Device fabrication was performed at ASRC, CUNY and CNF, Cornell.
\end{acknowledgments}

\end{document}